\documentclass[%
 reprint,
superscriptaddress,https://v2.overleaf.com/project/5bb52a0501ed504253226e97
showpacs,
amsmath,amssymb,
 aps,prl, 
]{revtex4-1}

\usepackage{graphicx}
\usepackage{dcolumn}
\usepackage{bm}
\usepackage{amsmath}
\usepackage{natbib}
\usepackage{color}
\usepackage[columnwise]{lineno}

\begin{document}

\preprint{APS/123-QED}

\title{Incorporating tunability into a universal scaling framework for shear thickening}

\author{Meera Ramaswamy}
\affiliation{Department of Physics, Cornell University, Ithaca, New York 14853, USA}
\author{Itay Griniasty}
\affiliation{Department of Physics, Cornell University, Ithaca, New York 14853, USA}
\author{James P Sethna}
\affiliation{Department of Physics, Cornell University, Ithaca, New York 14853, USA}
\author{Bulbul Chakraborty*}
\affiliation{Department of Physics, Brandeis University, Waltham, Massachusetts, USA}
\author{Itai Cohen*}
\affiliation{Department of Physics, Cornell University, Ithaca, New York 14853, USA}

\date{\today}

\begin{abstract}
Recently, we proposed a universal scaling framework that shows shear thickening in dense suspensions is governed by the crossover between two critical points: one associated with frictionless isotropic jamming and a second corresponding to frictional shear jamming. Here, we show that orthogonal perturbations to the flows, an effective method for tuning shear thickening, can also be folded into this universal scaling framework. Specifically, we show that the effect of adding in orthogonal shear perturbations (OSP) can be incorporated by simply altering the scaling variable to include a multiplicative term that decreases with the normalized OSP strain rate. These results demonstrate the broad applicability of our scaling framework, and illustrate how it can be modified to incorporate other complex flow fields.   
\end{abstract}

\pacs{Valid PACS appear here}

\maketitle

Shear thickening suspensions demonstrate an increase in the viscosity upon increasing the suspension stress \cite{morris2020shear, brown2010generality}. This increase in the viscosity has been attributed to the change in the nature of contacts from lubrication to frictional \cite{lin2015hydrodynamic, royer2016rheological, wyart2014discontinuous, seto2013discontinuous, mari2014shear} and can be very large and even lead to jamming at sufficiently high packing fractions. We recently proposed a scaling formulation that describes the shear thickening suspension viscosity as crossover between behaviours governing two different critical points. At small applied stresses, the suspension viscosity is determined by the distance to the frictionless isotropic jamming critical point. As the shear stress is increased and frictional contacts form between the particles, the flow properties transition to a new universal behavior, governed by the distance to the frictional shear jamming critical point, which occurs at lower volume fractions \cite{ramaswamy2021universal}. Interestingly, a growing body of work is showing that incorporating additional multi-directional flows can be used to modify the contact network and tune the suspension viscosity \cite{lin2016tunable,niu2020tunable,ness2018shaken,sehgal2019using,han2018shear,chacko2018shear, gillissen2019constitutive, gillissen2020constitutive}. Whether these more complex flows can be incorporated into our universal framework remains poorly understood.

\begin{figure}[t]
\includegraphics[width=\columnwidth]{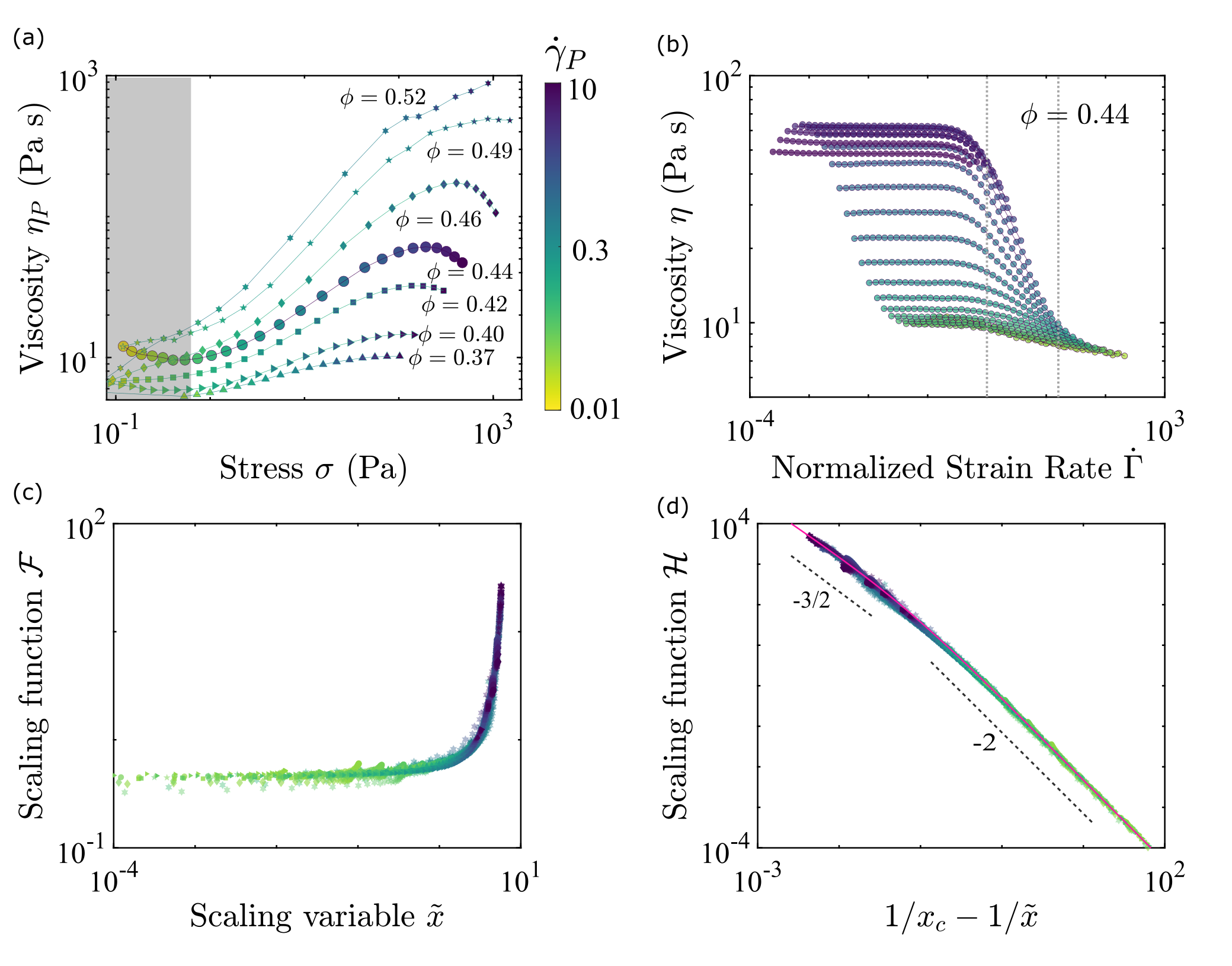}
\caption{\label{fig:FlowCurve} \textbf{Viscosity upon application of primary and orthogonal superposition shear and scaling collapse \textbf{(a)}}. The viscosity as a function of applied primary stress for suspensions of different volume fractions ranging from $\phi = 0.37$ to $\phi = 0.52$ with no OSP flow. The different colors correspond to different primary strain rates, ranging from 0.01 $s^{-1}$ (yellow) to 10 $s^{-1}$ (dark blue). The extent of shear thickening increases with increasing volume fraction. The shaded grey region are stresses at which suspensions undergo shear thinning, and will be not included in the scaling collapse. \textbf{(b)} The viscosity as a function of the normalized strain rate $\dot\Gamma = \gamma_{OSP}\omega_{OSP}/\dot\gamma_P$ for $\phi = 0.44$. A range of primary strain rates are applied ranging from 0.01 $s^{-1}$ (yellow) to 10 $s^{-1}$ (dark blue), and the OSP flow is varied by varying angular frequency $\omega_{OSP}$ from 0.05 $s^{-1}$ to 100 $s^{-1}$ at a constant strain amplitude $\gamma_{OSP} = 0.05$. We see that the decrease in the viscosity appears to occur at $\dot\Gamma \approx 1$ for all primary strain rates. \textbf{(c)} The scaling function $\mathcal{F} = \eta(\phi_0 - \phi)^2$ as a function of the  scaling variable, $\tilde x(\sigma, \dot\Gamma, \phi) = f(\sigma)g(\dot\Gamma)C(\phi)/(\phi_0 - \phi)$ where $f(\sigma) = e^{-(\sigma^*/\sigma)^{0.75}}$ and $g(\dot\Gamma) = e^{-\dot\Gamma/\dot\Gamma^*}$. We see excellent scaling collapse of the data across all stresses onto a single universal curve that diverges at $\tilde x = x_c$. \textbf{(d)} The scaling function $\mathcal{H} = \eta(\zeta(\sigma, \dot\Gamma, \phi))^2$ versus $|1/x_c - 1/\tilde x|$ where $\zeta(\sigma, \dot\Gamma, \phi) = f(\sigma)g(\dot\Gamma)C(\phi)$. Scaling the data in this manner clearly shows the beginning of the crossover between frictionless jamming $\tilde{x}^{-2}$ and frictional jamming $\sim \tilde{x}^{-1.5}$ (solid and dashed gray lines) discovered in our earlier work~\cite{ramaswamy2021universal}.}
\end{figure}

Here, we investigate how tunability via orthogonal superimposed perturbations (OSP) modifies the scaling relations. The basic idea is simple - by applying small amplitude sinusoidal flows orthogonal to the primary flow driving thickening, we can break up the force chains and alter the viscosity. Importantly, applying the scaling analysis to OSP flows requires enormous amounts of data collected over the broadest range of parameters possible. While previous OSP experiments clearly demonstrated the ability to tune the suspension viscosity, they were only conducted at one volume fraction and one shear rate \cite{lin2016tunable}. In this work, we perform experiments on a cornstarch in glycerol suspension for seven volume fractions ranging from $\phi = 0.37$ to $\phi = 0.52$ and for shear rates corresponding to the entire thickening transition. 

The average viscosity in the primary flow direction is measured with an ARES G2 rheometer, using a double wall Couette geometry. The primary flow is generated by the continuous rotation of the outer cup about the vertical z axis and the orthogonal flow is generated by the vertical oscillation of the inner bob. A steady primary flow rate ($\dot{\gamma}_P$) at sufficiently large imposed strain rates results in formation of force chains and thickening of the suspension. The resulting viscosity is tuned by superimposing a small amplitude, $\gamma_{OSP} \le 0.05$, sinusoidal orthogonal flow. 

The resulting measurements and the scaling collapse of the data are shown in Fig.~\ref{fig:FlowCurve}. In the absence of any orthogonal flows we observe typical thickening behavior (Fig.~\ref{fig:FlowCurve}\textbf{(a)}). Here, each volume fraction is indicated with a different symbol, and each primary strain rate is indicated with a different color (Fig.~\ref{fig:FlowCurve} side bar). The effect of OSP flow on the suspension viscosity at one volume fraction, $\phi=0.44$, is shown in Fig.~\ref{fig:FlowCurve}\textbf{(b)}. At each primary shear rate, OSP flow is applied with a fixed amplitude $\gamma_{OSP} = 0.05$, and angular frequencies, $\omega_{OSP}$, ranging from  $0.05 s^{-1}$ to 100 $s^{-1}$. As expected, at small frequencies there is little to no change in the viscosity, and at larger frequencies the suspension dethickens and the viscosity decreases. In addition, the viscosity in suspensions with OSP flows is expected to be governed by the ratio of the force chain disruption to force chain formation, which in its simplest form is captured by the dimensionless strain rate $\dot\Gamma = \gamma_{OSP}\omega_{OSP}/\dot\gamma_P$ \footnote{These cornstarch suspensions are at low Reynolds numbers. As such, there are no inherent time scales in the system, rather, just a strain scale required for the formation of force chains. Thus, force chain formation is proportional to the accumulated primary strain in a given interval, and force chain disruption is proportional to the number of orthogonal shear cycles that happen in that interval. The extent to which the suspension dethickens is then governed by the dimensionless variable  $\dot\Gamma = \gamma_{OSP}\omega_{OSP}/\dot\gamma_P$.} (Fig~\ref{fig:FlowCurve}b). Consistent with this formulation, we find that the viscosity starts decreasing at $\dot\Gamma \approx 1$ for all primary shear rates. The wide range of viscosities observed makes this OSP protocol an ideal first step to investigate the use of the scaling formulation in multi-directional flows. 

The scaling theory \cite{ramaswamy2021universal} in the absence of multi-directional flows uses the collapse of experimental data to write the viscosity as a function of a single scaling variable, $x_1 = \frac{f(\sigma)C(\phi)}{(\phi_0 - \phi)}$, that includes a stress dependent fraction of frictional contacts, $f(\sigma)$, a multiplicative factor $C(\phi)$ that depends on the volume fraction, $\phi$, and the distance to the frictionless isotropic jamming point $\phi_0 - \phi$.
\begin{equation}
    \eta(\phi_0 - \phi)^2 = \mathcal{F}(x_1)
\end{equation}
 At small $x_1$, $\mathcal{F}$ is constant and the viscosity diverges as $(\phi_0 - \phi)^{-2}$, consistent with universal scaling established for isotropic frictionless jamming. At larger values of $x_1$ this universal function $\mathcal{F}$ was found to have a divergence associated with frictional shear jamming. Here, the viscosity increased as $(x_c - x_1)^{-\delta}$, with $\delta \sim 3/2$ indicating that frictional shear jamming is associated with a different universality class. 
 
\begin{figure*}[t]
\includegraphics[width=0.9\linewidth]{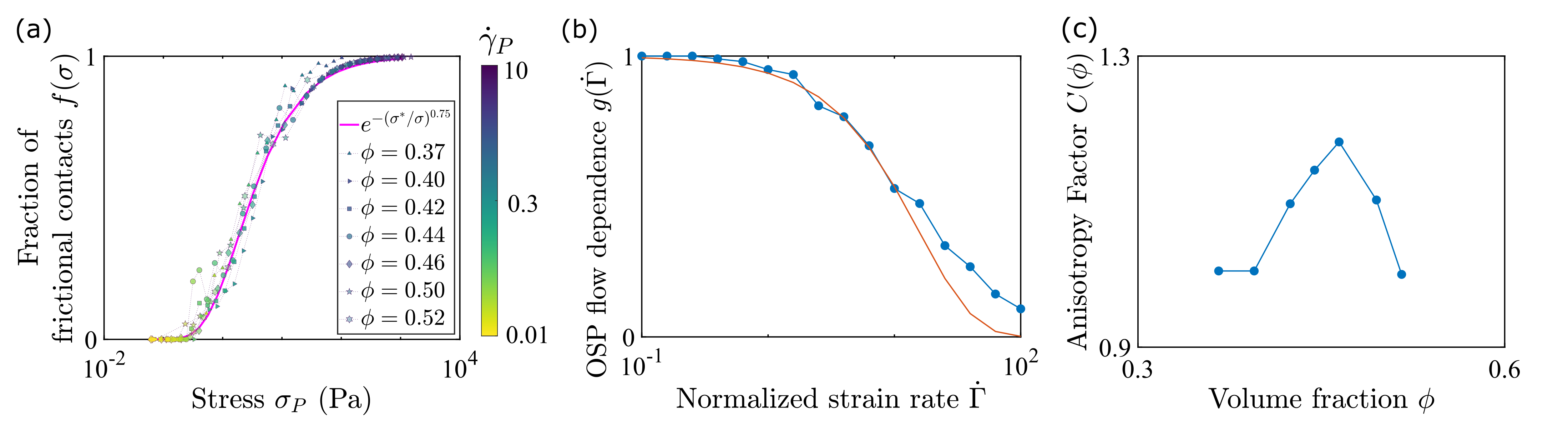}
\caption{\label{fig:FinalScalingFunctions} \textbf{Final functions in the scaling variable} \textbf{(a)} $f(\sigma)$ as a function of the applied stress. The symbols are the point by point values of $f(\sigma)$ for each volume fraction that best collapses the data.  The solid pink line is the stretched exponential form $f(\sigma) = e^{-(\sigma^*/\sigma)^{0.75}}$ that captures the point by point data well, and is used for the best scaling collapse of the data. \textbf{(b)} Multiplicative function of $\dot\Gamma$, $g(\dot\Gamma)$, used to collapse the data. The blue data is the point by point value for each $\dot\Gamma$ that best collapses the data. The red line is an exponential function of the normalized strain rate $e^{-\dot\Gamma/\dot\Gamma^*}$ that captures the transition in $g(\dot\Gamma)$ with increasing $\dot\Gamma$. \textbf{(c)}  The anisotropy factor $C(\phi)$ used for the scaling collapse of the data.}
\end{figure*}

Inspired by our previous work, we attempt a scaling collapse using a new scaling variable that contains a product of independent functions that depend on the stress $\sigma$, dimensionless strain rate, $\dot\Gamma$, and volume fraction, $\phi$. We find that this approach works extremely well collapsing our entire data set (Fig.~\ref{fig:FlowCurve}\textbf{(c)}). Similar to the previous analysis, we express the viscosity as
\begin{equation}
    \eta (\phi_0 -\phi)^2 = \mathcal{F}\left(\tilde x\right)
\end{equation}
but this time with 
\begin{equation}
    \tilde x (\sigma, \dot\Gamma, \phi) =\frac{\zeta(\sigma, \dot\Gamma, \phi)}{(\phi_0 - \phi)} = \frac{f(\sigma)g(\dot\Gamma)C(\phi)}{(\phi_0 - \phi)} 
    \label{eq:scaling}
\end{equation}
where the second equality captures a conjecture that $\zeta$ can be expressed as a product of three independent functions. The inclusion of the multiplicative term, $g(\dot\Gamma)$ is similar to those used for introducing additional constraints \cite{guy2018constraint} or a minimum strain \cite{han2018shear} required for thickening. Impressively, this approach enables scaling collapse of the data across all volume fractions, stresses and OSP flow rates and illustrates that our framework is indeed general enough to incorporate the effects of multi-directional flows. 

To determine the functions $C(\phi)$, $g(\dot\Gamma)$ and $f(\sigma)$, we use an iterative process. We start with our previous form for $f(\sigma) = e^{-\sigma^*/\sigma}$, where the stress used to calculate $f(\sigma)$ is always the primary stress, $\sigma = \eta_P *\dot\gamma_P$. Using the scaling collapse of $\dot\Gamma = 0$ data across different volume fractions, we estimate $C(\phi)$. With this function $C(\phi)$ and the initial $f(\sigma)$, we determine the values of $g(\dot\Gamma)$ that best collapses the entire data set. Next, holding $g(\dot\Gamma)$ and $C(\phi)$ fixed, we find the new values of $f(\sigma)$ that optimizes scaling collapse. This process of holding two of the functions $f(\sigma)$, $g(\dot\Gamma)$, and  $C(\phi)$, fixed to determine the third is repeated cyclically until a good scaling collapse of the data is achieved (see SI for intermediate scaling collapses and more details). The final functions are shown in Fig.~\ref{fig:FinalScalingFunctions}. We find that the best collapse is obtained when the stress dependence is approximated by a stretched exponential $f(\sigma) = e^{-(\sigma^*/\sigma)^{0.75}}$, the dependence on orthogonal shear is approximated by an exponential $g(\dot\Gamma) = e^{-\dot\Gamma/\dot\Gamma^*}$, and the dependence on volume fraction is non-monotonic (Fig.~\ref{fig:FinalScalingFunctions}).


To determine the critical exponents governing shear thickening in the presence of OSP flows, we conduct a Cardy scaling analysis \cite{cardy1996scaling,ramaswamy2021universal}, and plot the scaling function
\begin{equation}
    \zeta^2 \eta = \tilde{x}^2 \mathcal{F}(\tilde{x}) = \mathcal{H}\left(\frac{1}{x_c} - \frac{1}{\tilde x} \right)
\end{equation}
in Fig.~\ref{fig:FlowCurve}\textbf{(d)}. We find excellent scaling collapse over four orders of magnitude in the scaling variable and seven orders of magnitude in the universal scaling function, $\mathcal{H}$. At small values of $\tilde x$, the data scales with an exponent of -2, as expected for frictionless jamming. As $\tilde x \rightarrow x_c$, the data show a clear crossover to a different scaling exponent associated with frictional shear jamming, which is consistent with the -3/2 value reported previously \cite{ramaswamy2021universal} (see SI for more details).

\begin{figure}[t]
\includegraphics[width=\linewidth]{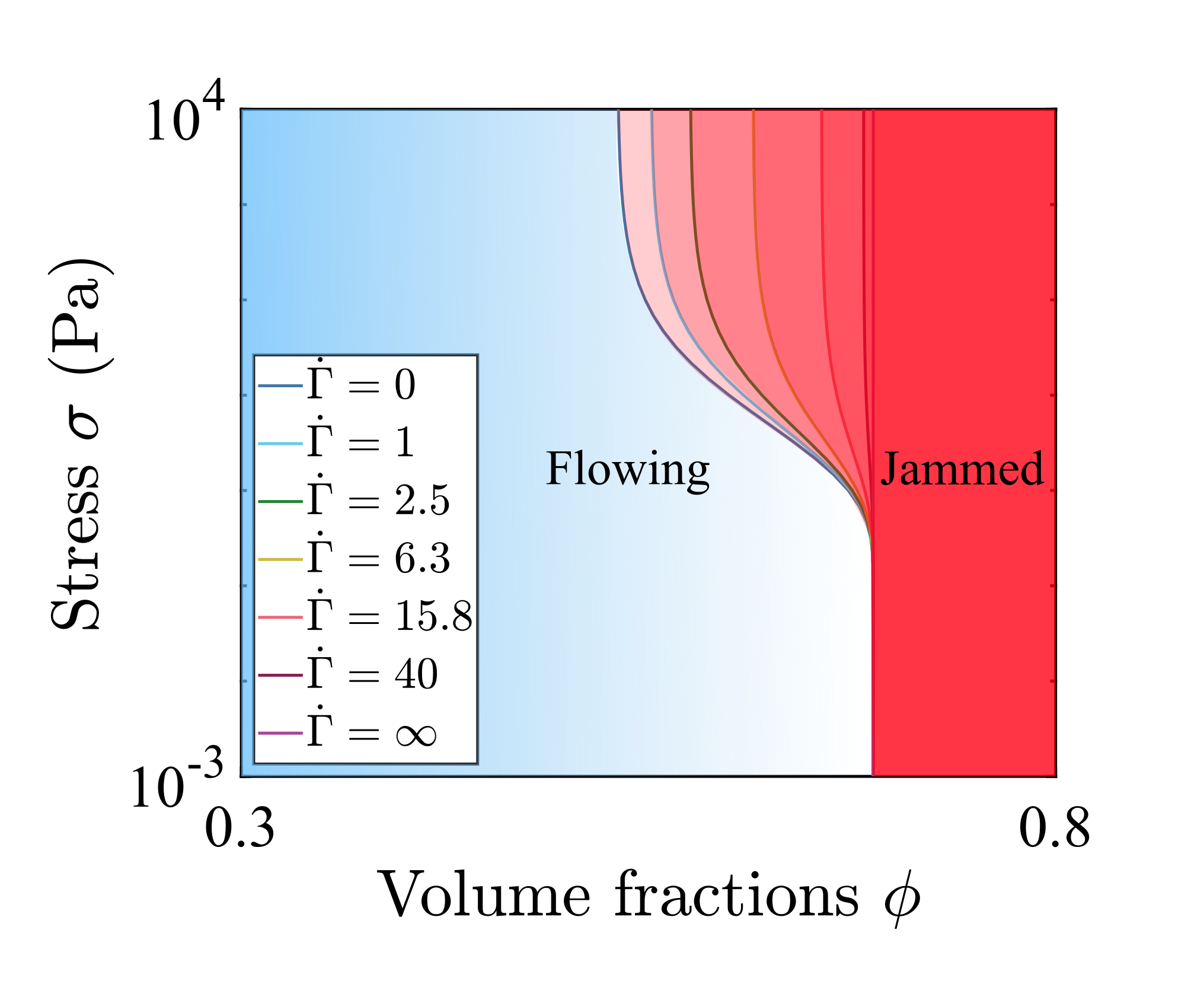}
\caption{\label{fig:PhaseDiagram} \textbf{Jamming phase diagram as derived from the scaling analysis}. The shear jamming lines for different normalized orthogonal strain rates $\dot{\Gamma}$, indicated by the different colours, are drawn in the $(\phi,\sigma)$ plane. The larger the normalized OSP the larger the region in which the suspension flows.
At small values of $\dot{\Gamma}$ the suspension is unaffected by orthogonal shear. At large $\dot{\Gamma}$ however, the formation of force chains is completely impeded, and the volume at which the suspension jams at high stress is equal to the isotropic jamming volume fraction at zero stress. }
\end{figure}

Using both the universal and the material specific parameters in the scaling collapse, we generate the evolving shear jamming boundary for this system. This surface is determined by the viscosity divergence at $x = x_c$ and is a function of all three parameters $\phi$, $\sigma$, and $\dot \Gamma$ (see SI for details). We visualize this surface by plotting the jamming volume fraction as function of the stress for different values of $\dot\Gamma$ in Fig.~\ref{fig:PhaseDiagram}. At small values of $\dot \Gamma$, the shear jamming volume fraction at large stresses is much smaller than that for isotropic frictionless jamming. As the OSP strain rate increases, the shear jamming line straightens, with the volume fraction for shear jamming approaching that for frictionless isotropic jamming as $\dot \Gamma \rightarrow \infty$, while retaining the critical exponent associated with frictional shear jamming at all non-zero applied stresses.

Similarly, we find that the discontinuous shear thickening region (shown in the SI) where $d\log \eta/d \log \sigma > 1$ shrinks with increasing OSP strain. Specifically, in the large $\tilde x$ limit where $\mathcal{F} \sim (x_c - \tilde x)^\delta$, this condition can be written as: 
\begin{equation}
    \frac{- \sigma \delta }{(x_c - \tilde x)}\frac{d f(\sigma)}{d \sigma} \frac{C(\phi) g(\dot\Gamma)}{(\phi_0 - \phi)} > 1.
\end{equation}
As $\dot \Gamma \rightarrow \infty$, i.e. as $g(\dot\Gamma) \rightarrow 0$, this relationship only holds when $\tilde x \rightarrow x_c$ or equivalently when $\phi \rightarrow \phi_0$. Thus, as the OSP strain rate is increased the boundary for the discontinuous shear thickening region approaches the shear jamming line. 

The phase diagram in Fig.~\ref{fig:PhaseDiagram} illustrates how orthogonal shear flows can dramatically alter the suspension properties by driving the system from thickening behaviour governed by the frictional shear jamming critical point towards behaviour governed purely by the frictionless isotropic critical point.  As such, OSP flows play a role reciprocal to that played by the shear stress, which increases frictional contacts and drives the system away from frictionless isotropic jamming critical point. This reciprocal behaviour and the striking similarity between the functional forms for $g(\dot\Gamma)$ and $f(\sigma)$, suggests that their product, $f(\sigma)g(\dot\Gamma)$, dictates the fraction of frictional contacts in the system. Whether additional strategies for manipulating the suspension, such as acoustics or other time dependent flows, can be treated similarly and folded into this scaling analysis remains an exciting avenue for future research. 

Importantly, this new understanding, revealed through the scaling analysis, enables unprecedented control over accessing various regions in the phase diagram. For instance, by switching the orthogonal shear flows on we can drive the system into a high stress flowing state (pink region in Fig.~\ref{fig:PhaseDiagram}). Once these OSP flows are turned off, the system is forced into a previously inaccessible state deep in the shear jammed region. A similar approach can be used to prepare the system deep within the unstable discontinuous shear thickening region. As such, this framework opens novel avenues for generating unique material states. 

\section*{Acknowledgements}
We thank Edward Y. X. Ong, Danilo Liarte, Emanuela Del Gado, Eleni Katifori, Eric M. Schwen, and Stephen J. Thornton for valuable discussions, and Anton Paar for use of the MCR 702 rheometer through their VIP academic research program. For this work, IC, IG, and MR were generously supported by  NSF CBET award numbers: 2010118, 1804963, 1509308, and an NSF DMR award number: 1507607. BC was supported by NSF CBET award number 1916877 and NSF DMR award number 2026834. JPS was supported by NSF CBET-2010118.

\section*{Author Contributions}
MR is the corresponding author. BC and IC contributed equally to this work. All the authors were involved in discussions that shaped the idea and experiments in this manuscript. MR ran the experiments. MR, BC, IC, and JPS analyzed the experimental data. MR, BC, JPS and, IC wrote the manuscript with all authors contributing. 

\bibliography{references}
\bibliographystyle{naturemag}
\end{document}


\preprint{APS/123-QED}

\title{Incorporating tunability of shear thickening transitions into a universal scaling framework}
\renewcommand{\theequation}{S\arabic{equation}}
\renewcommand{\thefigure}{S\arabic{figure}}
\renewcommand{\thesection}{S\arabic{section}}

\begin{center}
   \large{\textbf{Incorporating tunability of shear thickening transitions into a universal scaling framework}}
    
    \textbf{Supplementary Materials}
\end{center}

\section{\label{Methods}Sample Preparation and Experimental Protocol}
The samples were prepared by weighing out the cornstarch (Argo) and glycerol (Sigma-Aldrich). We used cornstarch suspensions as they are ubiquitous in the study of shear thickening suspensions. Glycerol is used as a solvent as it is less easily absorbed by the cornstarch and can be easily loaded into the small gap in the double gap couette geometry. The volume fraction was calculated as 
\begin{equation}
    \phi = \frac{\rho_g \phi_M}{\rho_g \phi_M + (1- \phi_M)\rho_s}
\end{equation}
where $\phi$ is the volume fraction, $\phi_M$ is the mass fraction of cornstarch, $\rho_g$ is the density of glycerol, and $\rho_s = 1.62$g/cm$^3$ is the density of the cornstarch.  The suspension is mixed manually with a spatula for 10 minutes and then used immediately. 

An TA instruments, ARES-G2 strain controlled rheometer is used with a double wall concentric cylinder geometry and the OSP flow. The sample was presheared at a constant strain rate of 1$s^{-1}$ for five minutes. The sample was then sheared with only a primary shear at shear rates ranging from $0.01s^{-1}$ to $10s^{-1}$. For the higher volume fractions, the strain rate ramp was stopped when the stress exceeded the maximum stress allowed by the rheometer. Following the steady shear ramp, the orthogonal oscillatory shear is applied. At each primary shear rate, orthogonal shear is applied with a constant strain amplitude of $0.05\%$, and at angular frequencies from $0.05 s^{-1}$ to 100 $s^{-1}$. The viscosity, measured as the stress along the axial direction divided by the primary shear rate is reported in the paper. 

\section{\label{sec:Phi0Calc} Estimation of $\phi_0$}
 \begin{figure}
 \begin{center}
\includegraphics[width=0.8\linewidth]{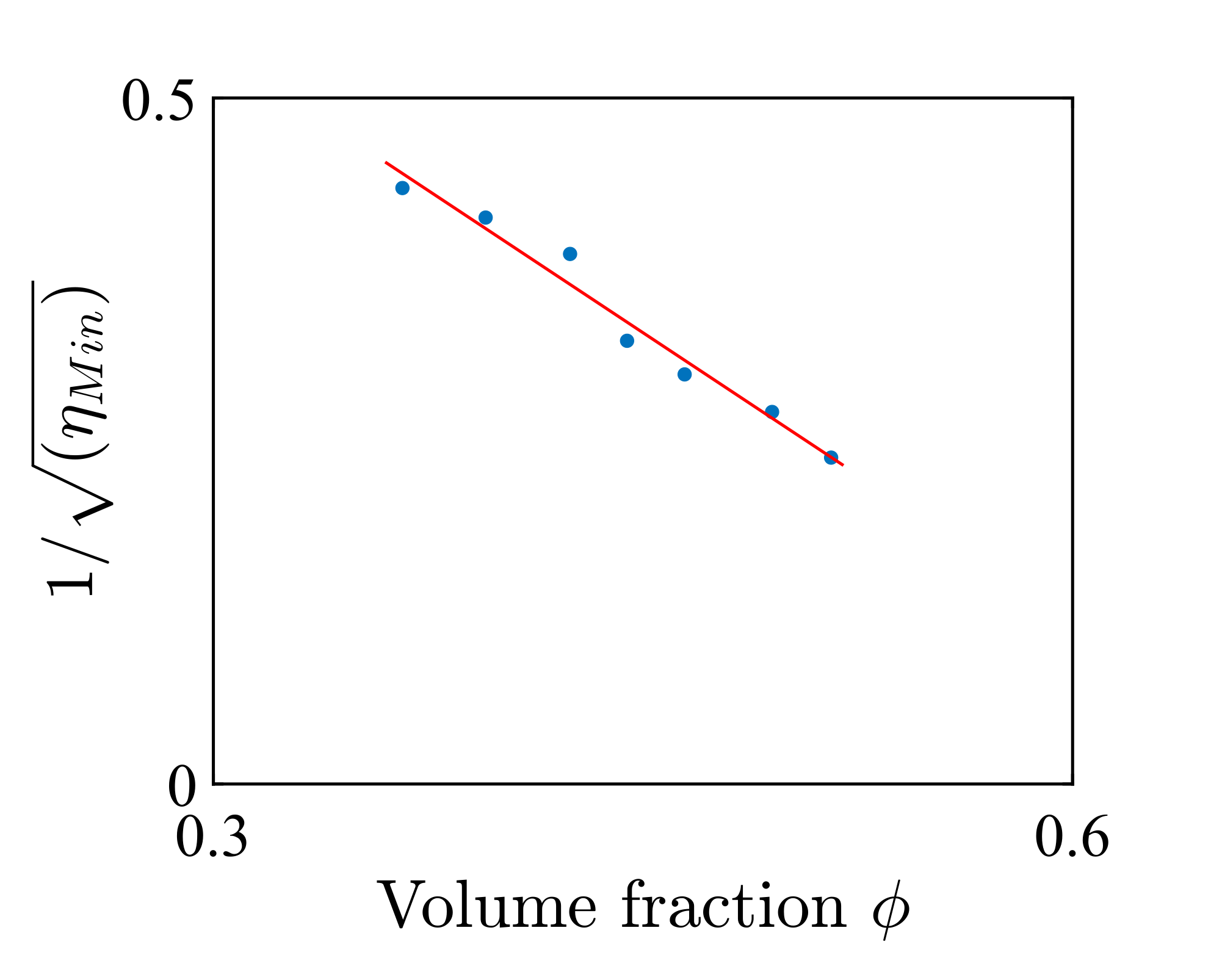}
\caption{\textbf{Estimation of $\phi_0$.} Plot of $\eta_{min}^{-1/2}$ as a function of the volume fraction $\phi$, where $\eta_{min}$ is the viscosity prior to shear thickening for the steady shear system. The red line is a linear fit to the data and the x intercept is the estimated value of $\phi_0$.}
\label{fig:Phi0Calc}
\end{center}
\end{figure}

We use the low strain rate viscosity with the primary shear rate alone across all volume fractions $\eta_{{Min}}$ to estimate the value of $\phi_0$. It has been shown previously that the low stress viscosity prior to shear thickening diverges at $\phi_0$ as $\eta \sim 1/(\phi_0 - \phi)^2$. Thus, we plot $1/\sqrt{\eta_{Min}}$ as a function of $\phi$, fit it to a straight line. The x intercept of the line is is used as  $\phi_0$ (Fig.~\ref{fig:Phi0Calc}). We find that this value of $\phi_0$ collapses the data well in the regime of small values of the scaling variable $\tilde x$.

\section{Scaling collapse}
\begin{figure}[t]
\includegraphics[width=\columnwidth]{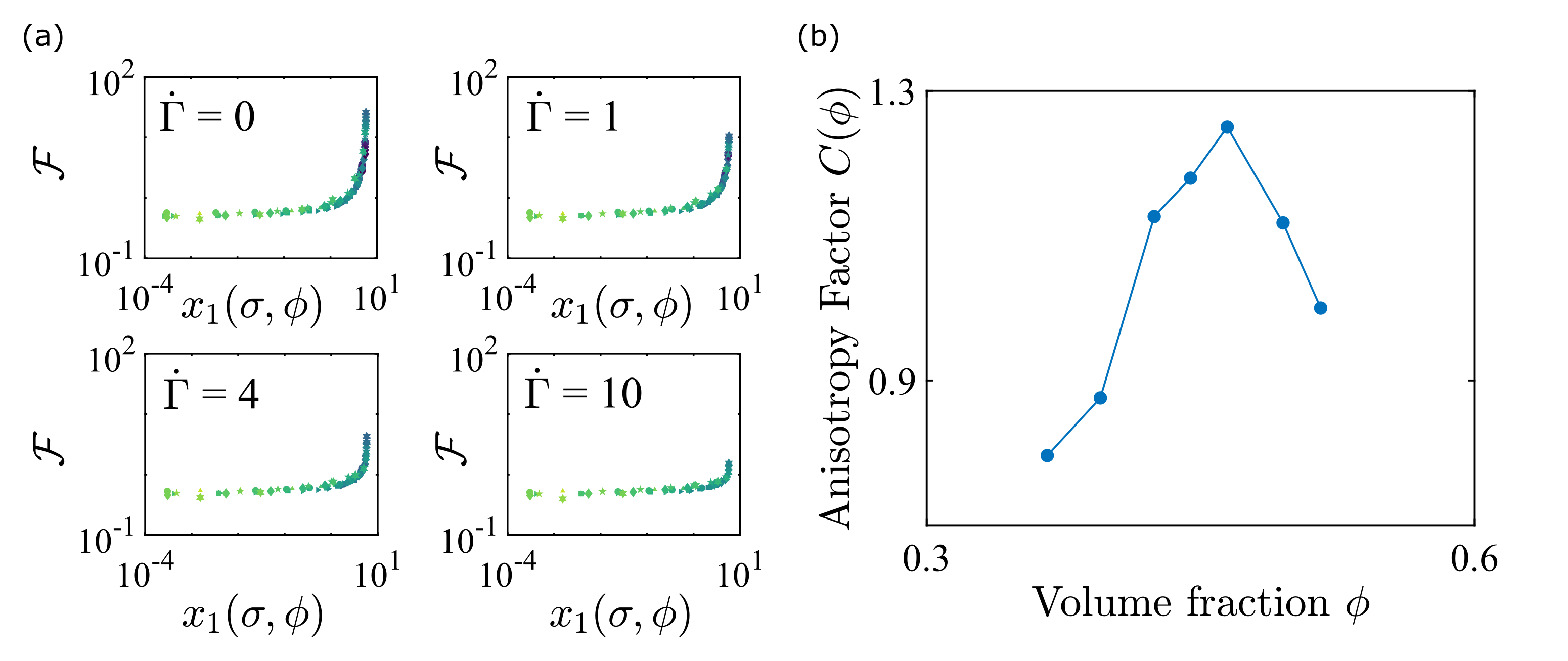}
\caption{\label{fig:FlowCurve} \textbf{Scaling collapse of data binned at constant $\dot\Gamma$} \textbf{(a)} Independent scaling collapse of data at each value of $\dot\Gamma$, following the protocol in \cite{ramaswamy2021universal}. The scaling variable, $x_1(\sigma, \phi) = e^{-\sigma^*/\sigma}C(\phi)/(\phi_0 - \phi)$, can be used to get excellent scaling collapse of the data to generate the scaling function, $\mathcal{F} = \eta(\phi_0 - \phi)^2$ for all values of $\dot\Gamma$. \textbf{(b)} The anisotropy factor $C(\phi)$ used for the scaling collapse of the data.}
\end{figure}

\begin{figure*}[t]
\includegraphics[width=0.9\linewidth]{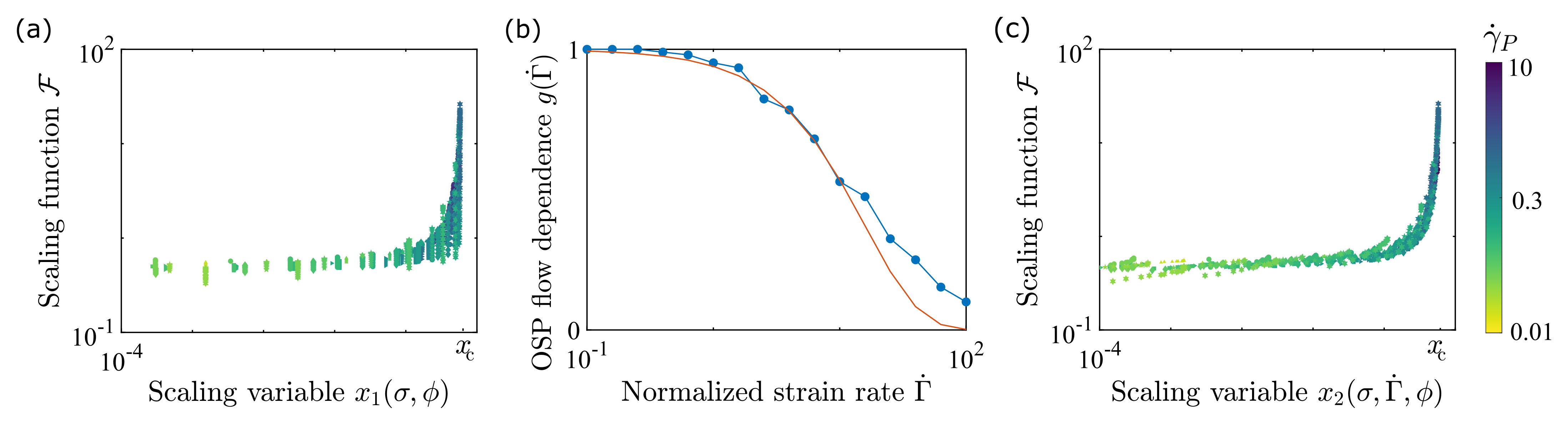}
\caption{\label{fig:ScalingConstGamma} \textbf{Scaling collapse of data with an exponential $f(\sigma)$} \textbf{(a)}. Failed scaling of data across all volume fractions, $\phi$, primary strain rates, $\dot\gamma_P$ and normalized strain rates $\dot\Gamma$, using previously determined scaling variable \cite{ramaswamy2021universal} that is independent of $\dot\Gamma$, $x_1(\sigma, \phi) = e^{-\sigma^*/\sigma}C(\phi)/(\phi_0 - \phi)$. The scaling function is $\mathcal{F} = \eta(\phi_0 - \phi)^2$. As expected, data at a given stress and volume fraction now has the same value of the scaling variable, irrespective of $\dot\Gamma$, and as such, the dethickening in the viscosity appears to occur at a constant $x_1$, evident in the lack of scaling collapse of the data. \textbf{(b)} Multiplicative function of $\dot\Gamma$, $g(\dot\Gamma)$, used to better collapse the data. The blue data is the point by point data that best collapses the data. The red line is an exponential function of the normalized strain rate $e^{-\dot\Gamma/\dot\Gamma^*}$ that captures the transition in $g(\dot\Gamma)$ with increasing $\dot\Gamma$. \textbf{(c)} Attempted scaling collapse with a new scaling variable, $x_2(\sigma, \dot\Gamma, \phi) = e^{-\sigma^*/\sigma}g(\dot\Gamma)C(\phi)/(\phi_0 - \phi)$, that depends on $\dot\Gamma$. We find that while the scaling collapse is better than in Fig~\ref{fig:ScalingConstGamma}\textbf{(a)}, the collapse is not perfect. In particular, we note that the data at different primary strain rates diverge at different values of the scaling variable $x_2(\sigma, \dot\Gamma, \phi)$}
\end{figure*}

 \begin{figure}
 \begin{center}
\includegraphics[width=0.7\linewidth]{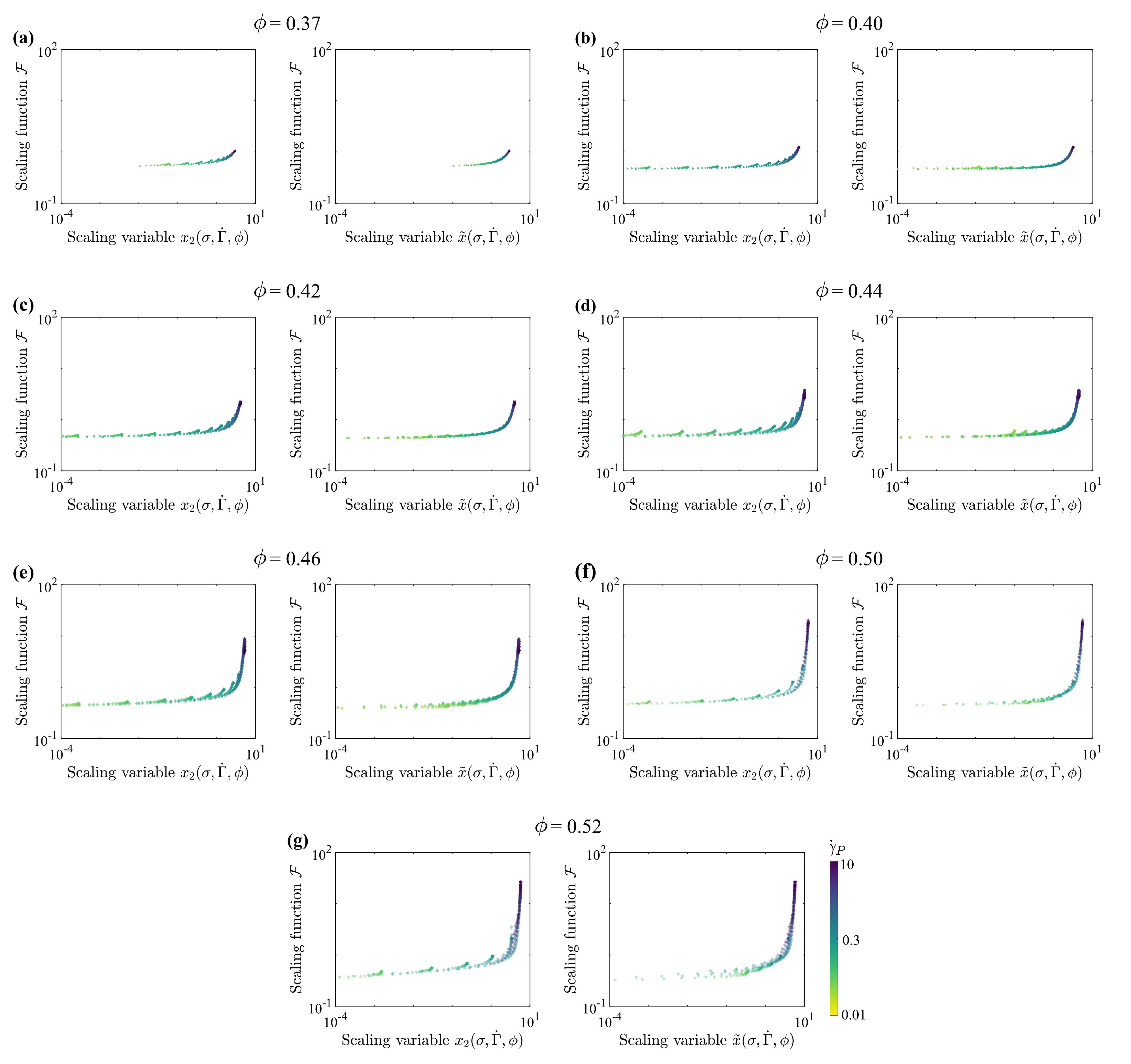}
\caption{\textbf{Scaling collapse of viscosity data at each volume fraction} The figures on the left show the failed scaling where the  scaling variable is $x_2(\sigma, \phi, \dot\Gamma) = C(\phi)e^{-\sigma^*/\sigma}e^{-\dot\Gamma/\dot\Gamma^*}/(\phi_0 - \phi)$ and the scaling function $\mathcal{F} = \eta(\phi_0 - \phi)^2$. We can clearly see that the data at different primary stresses diverge at different points, indicating that a stress dependent modification is required to the scaling variable. The figures on the right show the excellent scaling collapse obtained when $f(\sigma)$ was changed to a stretched exponential, where $\tilde x(\sigma, \phi, \dot\Gamma) = C(\phi)e^{-(\sigma^*/\sigma)^{0.75}}e^{-\dot\Gamma/\dot\Gamma^*}/(\phi_0 - \phi)$ and the scaling function $\mathcal{F} = \eta(\phi_0 - \phi)^2$. The different colours indicate the different stresses, with blue as the largest primary stress and green is the smallest primary stress.}
\label{fig:FailedScalingAllPhi}
\end{center}
\end{figure}

We first fit for the global parameter $\sigma^*$ by fitting the raw viscosity data to the Wyart and Cates model. Using this value of $\sigma^*$, we can calculate $f(\sigma) = e^{-\sigma^*/\sigma}$ for all packing fractions and stresses. Importantly, the stress in this expression is the stress in the \textit{absence} of any orthogonal flow. Thus the stress used in the scaling collapse is $\sigma = \eta_P \dot\gamma_P$. 

To collapse the data, we first attempt a scaling collapse for the steady state data $\dot\Gamma=0$. A volume fraction dependent function, $C(\phi)$, (Fig.~\ref{fig:FlowCurve}\textbf{(b)}) is then determined that best collapses the data across different volume fractions onto a single curve (Fig.~\ref{fig:FlowCurve}\textbf{(a)} upper left panel). As expected, we find excellent scaling collapse across several decades in the suspension viscosity and the scaling variable. We then bin the data into bins of constant $\dot\Gamma$. We find similarly good collapse at other OSP strain rates and the scaled curves for $\dot\Gamma = 1,4,$ and $10$ are shown in the remaining panels of Fig.~\ref{fig:FlowCurve}\textbf{(a)} using the same values of $C(\phi)$ and the same function $f(\sigma)$.   

Though similar, the curves for different values of $\dot\Gamma$ do not collapse (Fig.~\ref{fig:ScalingConstGamma}\textbf{(a)}), indicating that a scaling variable that depends only on $\sigma$ and $\phi$ is not sufficient for capturing the dethickening behavior. We thus extend the scaling variable to account for the decrease in the viscosity to OSP flow. To construct $g(\dot\Gamma)$, we manually shift the curves for each $\dot\Gamma$ onto the $\dot\Gamma = 0$ curve. We plot the shift, $g(\dot\Gamma)$ versus $\dot\Gamma$ in Fig.~\ref{fig:ScalingConstGamma}\textbf{(b)} (blue points). We find that the behaviour of $g(\dot\Gamma)$ is well captured by the functional form $g(\dot\Gamma) = e^{-\dot\Gamma/16}$. Incorporating this functional form into $x_2$, 
\begin{equation}
    x_2(\sigma, \phi, \dot\Gamma) = \frac{C(\phi)e^{-\sigma^*/\sigma}e^{-\dot\Gamma/\dot\Gamma^*}}{(\phi_0 - \phi)}
\end{equation}
we attempt a scaling collapse for the entire data set (Fig.~\ref{fig:ScalingConstGamma}\textbf{(c)}). While much improved, we find that data at different primary stresses diverge at different values of $x_2$. This deviation is evident especially when the scaling collapse is attempted with each volume fraction independently (Fig.~\ref{fig:FailedScalingAllPhi}). 

The systematic stress dependent deviation in the scaling collapse suggests that our form for $f(\sigma)$ should be modified. To determine this modified form, we shift the curves for each volume fraction and applied stress onto a single curve. We find that there is no dependence on the volume fraction and that the entire range is well described by a stretched exponential of the form, $f(\sigma) =- e^{-(\sigma^*/\sigma)^{0.75}}$. This stretched exponential form results in excellent scaling collapse for data at each volume fraction (Fig.~\ref{fig:FailedScalingAllPhi}). Using this modified form for $f(\sigma)$, we determine again the best value of  $C(\phi)$ that collapses the data (see main text for the final $C(\phi)$ and $f(\sigma)$). With this new form of $f(\sigma)$ and $C(\phi)$ and a scaling variable
\begin{equation}
    \tilde x (\sigma, \dot\Gamma, \phi) = \frac{f(\sigma)g(\dot\Gamma)C(\phi)}{(\phi_0 - \phi)} \equiv \frac{\zeta(\sigma, \dot\Gamma, \phi)}{(\phi_0 - \phi)}
    \label{eq:scaling}
\end{equation}
we find excellent scaling collapse across all the volume fractions, strain rates and OSP flows. 

\section{\label{x_cx} Determining the scaling exponent}
 \begin{figure}
 \begin{center}
\includegraphics[width=0.6\linewidth]{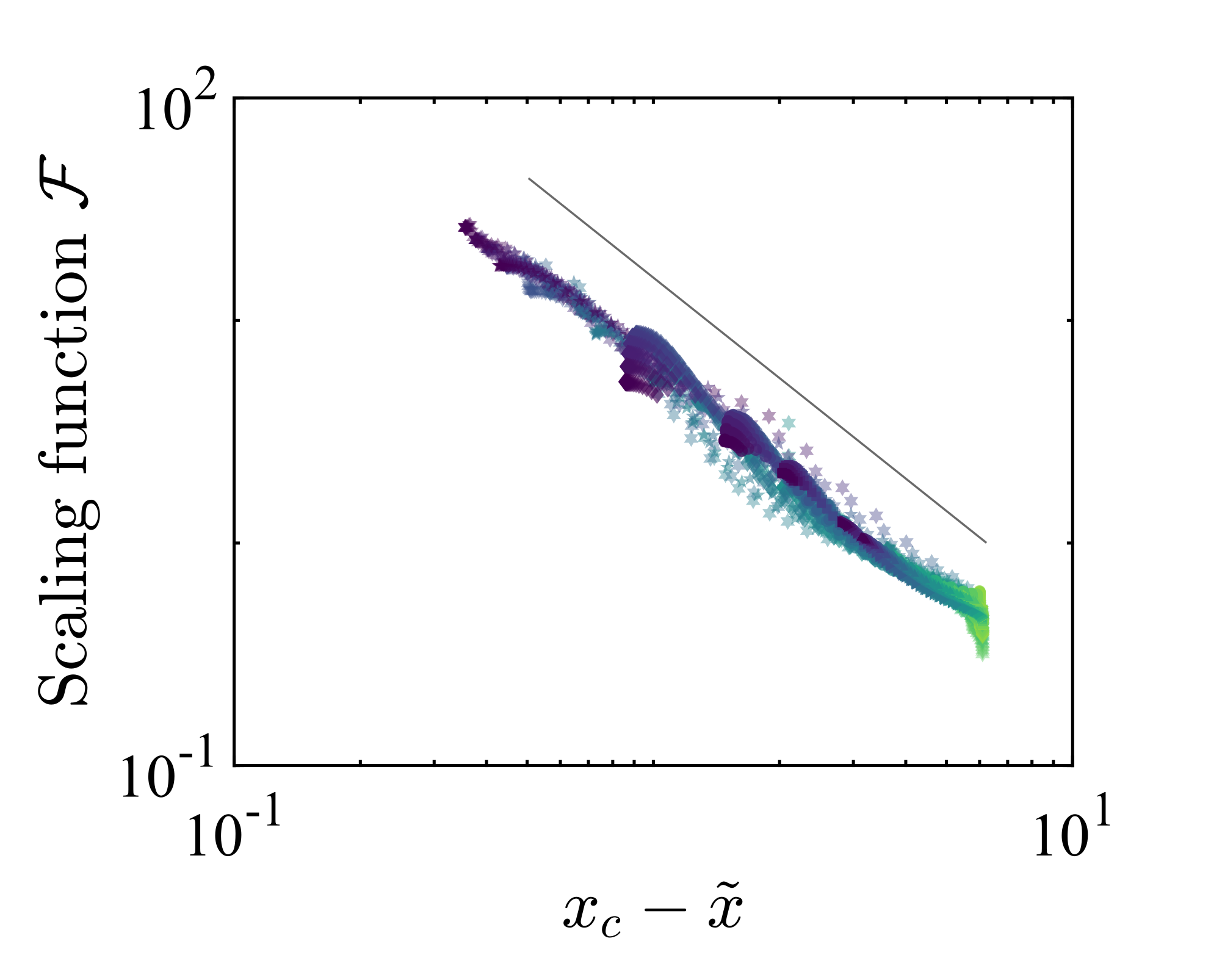}
\caption{\textbf{Power law scaling of cornstarch viscosity data.} The scaling function $\mathcal{F} = \eta(\phi_0 - \phi)^2$ as a function of the $x_c - \tilde x$ where the scaling variable $\tilde x = e^{-(\sigma^*/\sigma)^{0.75}}e^{-\dot\Gamma/\dot\Gamma_0}C(\phi)/(\phi_0 - \phi)$. We find that the data shows power law scaling with an exponent of $\sim -1.5$. The black solid line is a line with power law -1.5.}
\label{fig:x_c_x}
\end{center}
\end{figure}
From the scaling collapse of the data shown in Fig, 3b, the scaling function $\mathcal{F}$ appears to diverge at $x_c$: $\mathcal{F} \sim 1/(x_c - \tilde x)^\delta$. To determine the scaling exponent $\delta$, we plot the scaling function $\mathcal{F} = \eta(\phi_0 - \phi)^2$ as a function of $x_c - \tilde x$ (Fig.~\ref{fig:x_c_x}). We find that the divergence is indeed a power law, with $\delta \neq 2$. The best estimate for $\delta \approx -3/2$. However, since the data is still reasonably far from $x_c$, this value of $\delta$ should not be taken as a fixed parameter.

\section{\label{SJLime} Shear Jamming Line Calculation}
The shear jamming line is given by the divergence of the scaling function $\mathcal{F}$ at $\tilde x = x_c$. The scaling variable depends on the stress, volume fraction and the OSP flow rate: 
\begin{equation}
   \tilde x = \frac{f(\sigma) C(\phi)g(\dot\Gamma)}{(\phi_0 - \phi)} 
\end{equation}
where $f(\sigma) = e^{-(\sigma^*/\sigma)^{0.75}}$ and $g(\dot\Gamma) = e^{-\dot\Gamma/\dot\Gamma_0}$
Since $C(\phi)$ is fit for each volume fraction, we estimate that value of the $C(\phi)$ at high volume fraction by fitting the data points at the highest three volume fractions to a straight line: 

\begin{equation}
 C(\phi) = c_1 \phi + c_2   
\end{equation}

We can then determine the shear jamming volume fraction for each $\sigma$ and $\dot\Gamma$ as:

\begin{equation}
    \phi_{J, \sigma, \phi} = \frac{x_c \phi_0 - c_2 g(\dot\Gamma)f(\sigma)}{c_1 f(\sigma) g(\dot\Gamma) + x_c}
\end{equation}

For the given value of $C(\phi)$, $x_c = 6.1$. The values of $c_1$ and $c_2$ used to generate the phase diagram are listed in Table 1.

\begin{figure}[t]
\begin{center}
\includegraphics[width=0.6\linewidth]{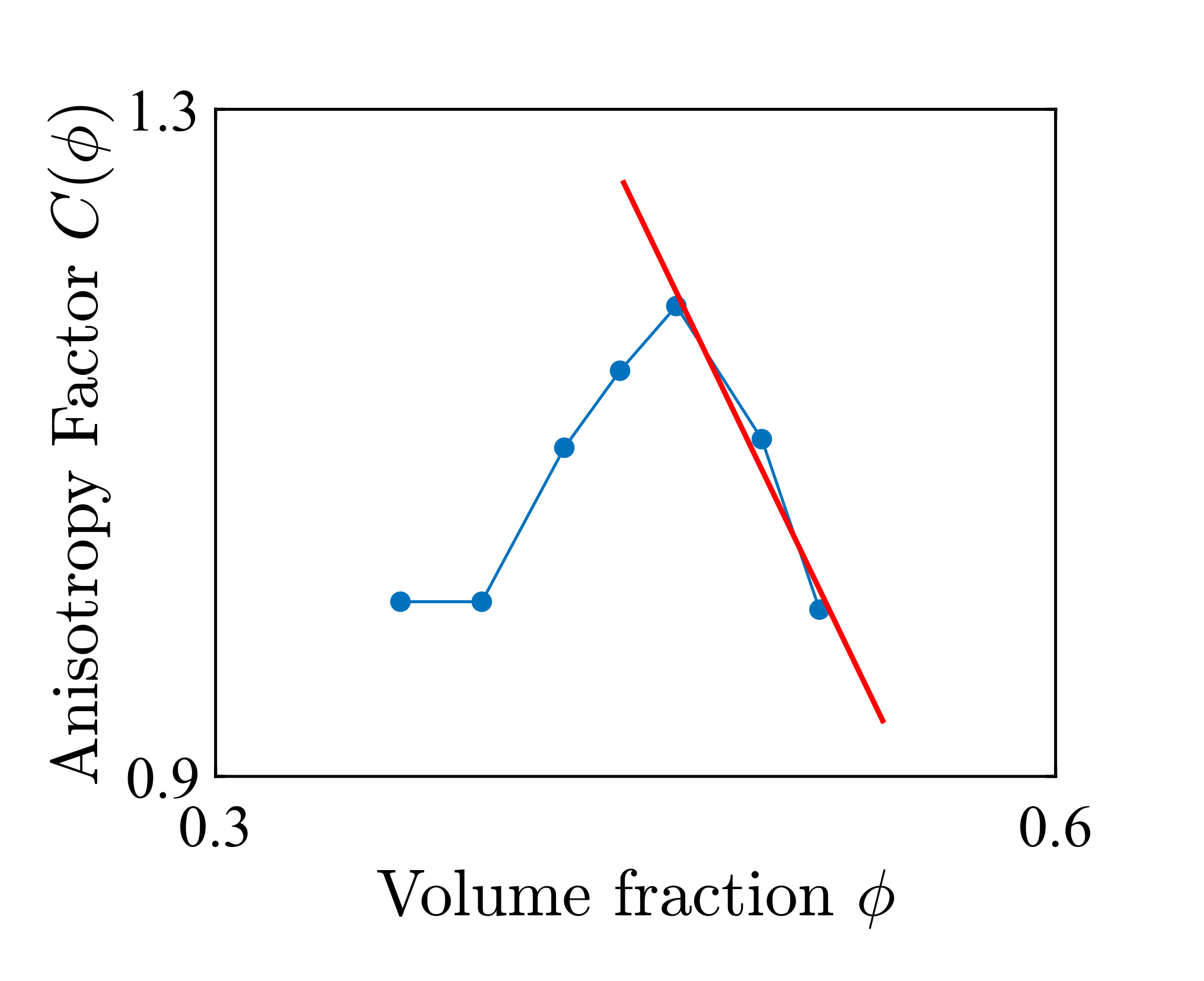}
\caption{\label{fig:CPhiFit} \textbf{Fits to the anisotropy $C(\phi)$}. The anisotropy factor $C(\phi)$ and the linear fit at high packing fraction (red) used to generate the phase diagram.}
\end{center}
\end{figure}

\section{\label{DSTLime}Estimating the effect of the OSP flow on the DST region:}
The DST line is calculated as 
\begin{equation}
    \frac{d \log \eta}{d \log \sigma} = 1
\end{equation}
or, 
\begin{equation}
    \frac{\sigma}{\eta}\frac{d\eta}{d\sigma} = 1
\end{equation}. 

Using the scaling form of the viscosity - 
\begin{equation}
    \eta (\phi_0 - \phi)^2 = \mathcal{F}\left(\tilde x\right)
\end{equation}, 
where 
\begin{equation}
    \tilde x = \frac{f(\sigma)g(\dot\Gamma)C(\phi)}{\phi_0 - \phi}
\end{equation}

Taking derivatives, 
\begin{equation}
    \frac{\sigma}{\eta}\frac{d\eta}{d\sigma}  = \frac{\sigma}{\eta}\frac{1}{(\phi_0 - \phi)^2}  \frac{d\mathcal{F}}{d \tilde x} \frac{d f(\sigma)}{d\sigma} \frac{C(\phi)g(\dot\Gamma)}{(\phi_0 - \phi)}= 1
\end{equation}

or
\begin{equation}
    \frac{\sigma}{\frac{\mathcal{F}}{(\phi_0 - \phi)^2}} \frac{1}{(\phi_0 - \phi)^2}\frac{d\mathcal{F}}{d \tilde x} \frac{d f(\sigma)}{d\sigma} \frac{C(\phi)g(\dot\Gamma)}{(\phi_0 - \phi)}= 1
\end{equation}

\begin{equation}
    \frac{\sigma}{\mathcal{F}}\frac{d\mathcal{F}}{d \tilde x} \frac{d f(\sigma)}{d\sigma} \frac{C(\phi)g(\dot\Gamma)}{(\phi_0 - \phi)}= 1
\end{equation}

Fig~\ref{fig:x_c_x} suggests that we can estimate $\mathcal{F}(\tilde x) = (x_c - \tilde x)^\delta$. In the limit that $g(\dot\Gamma) \rightarrow 0$, the left hand side of the equation

\begin{equation}
    \sigma \frac{1}{x_c} \frac{d f(\sigma)}{d\sigma} \frac{C(\phi)g(\dot\Gamma)}{(\phi_0 - \phi)} \rightarrow 0 
\end{equation}

As such, we expect that as the OSP flow increases and $g(\dot\Gamma)$ decreases, the DST region gets progressively smaller and vanishes. This behaviour is further reinforced by the calculations performed with the estimate for the crossover scaling function $\mathcal{H}$. 

\section{Comparisons with previous scaling collapse}
\begin{figure}[t]
\begin{center}
\includegraphics[width=0.6\linewidth]{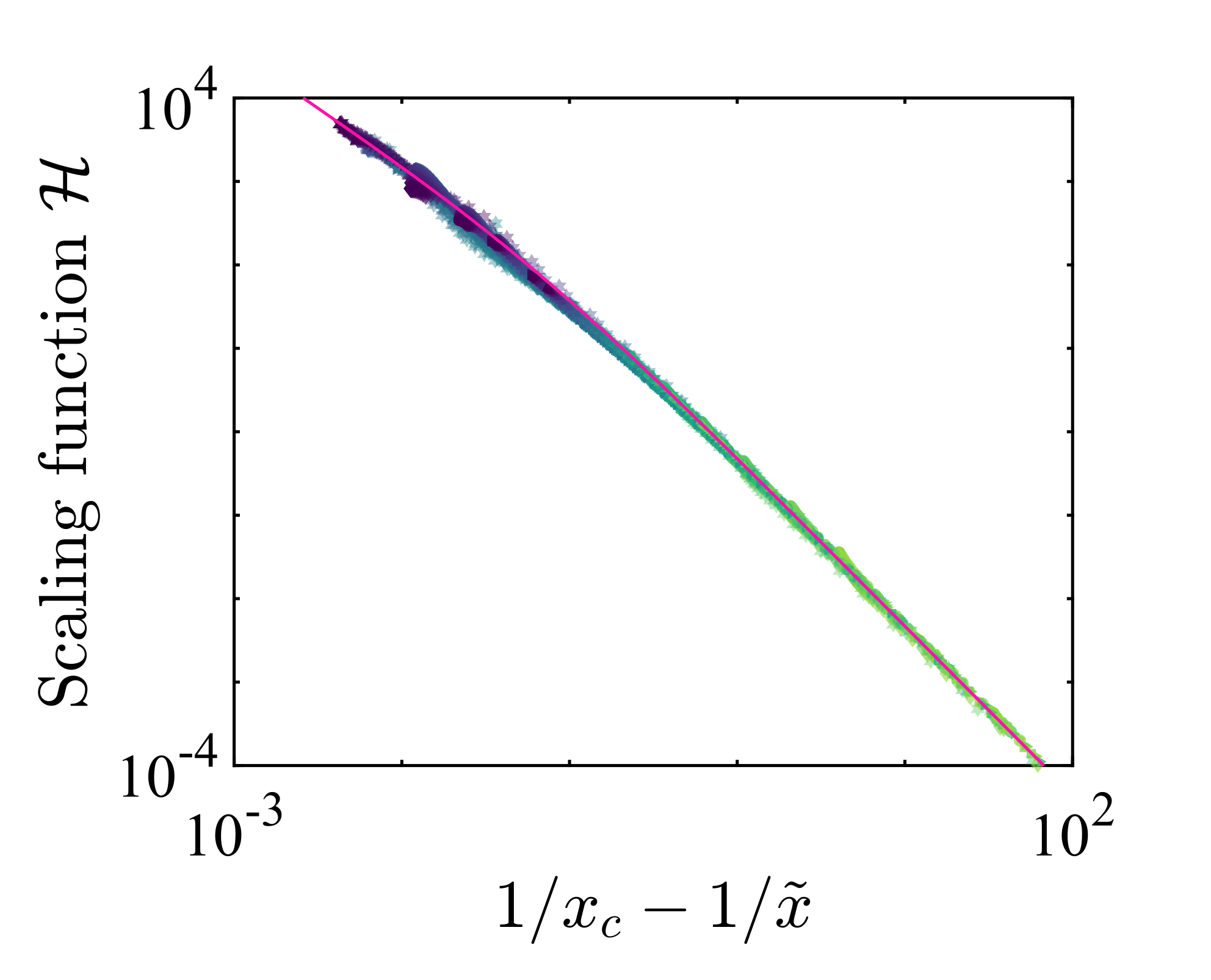}
\caption{\label{fig:CardyWithFit} \textbf{Cardy Scaling with fit from previous scaling function}. The scaling function $\mathcal{H} = \eta(\zeta(\sigma, \dot\Gamma, \phi))^2$ versus $|1/x_c - 1/\tilde x|$ where $\zeta(\sigma, \dot\Gamma, \phi) = f(\sigma)g(\dot\Gamma)C(\phi)$ as in Fig. 3\textbf{(c)}. The overlaid pink line is the scaling function from \cite{ramaswamy2021universal}, scaled by 1.7 to match the y axis and matching $x_c$.}
\end{center}
\end{figure}
The scaling collapse in Fig. 3\textbf{(c)} demonstrates a change in the exponent, however, we don't have data at sufficiently high stresses and volume fractions to fully determine the scaling collapse. However, we anticipate that the scaling function to be universal suggesting that the we should be able to overlay the scaling function obtained from the steady shear data in \cite{ramaswamy2021universal}. We overlay the crossover between two power laws fit to $\mathcal{H}$ onto the Cardy scaling with the OSP flow. To overlay the two curves, we match the values of $x_c$ by multiplying $C(\phi)$ by a constant. In addition we find that we need to multiply the fit by a constant factor of 1.7 to best overlap the two functions. We find that the scaling functions obtained with the scaling collapse of the OSP flow data is consistent with the scaling function obtained in the steady shear data in \cite{ramaswamy2021universal}. 

We can use the previous functional form of $\mathcal{H}$ to explore in more detail how the DST line changes with $g(\dot\Gamma)$. We determine the derivative the DST line as 
\begin{equation}
    \frac{d \log \eta}{d \log \sigma} = 1
\end{equation}
or, 
\begin{equation}
    \frac{\sigma}{\eta}\frac{d\eta}{d\sigma} = 1
\end{equation}

From the Cardy scaling collapse, we can write: 
\begin{equation}
    \eta(C(\phi)f(\sigma)g(\dot\Gamma))^2 = \mathcal{H}(1/\tilde x - 1/x_c)
\end{equation}

Taking the derivative of the viscosity, 
\begin{equation}
    \frac{\sigma}{\eta}\frac{d\eta}{d\sigma} = \frac{\sigma}{\eta}\frac{1}{(C(\phi)g(\dot\Gamma))^2}\left( \frac{-2}{f^3}\mathcal{H}\frac{df}{d\sigma} + \frac{1}{f^2}\frac{d\mathcal{H}}{d\sigma}\right) = 1
\end{equation}

Noting that $\mathcal{H}$ is a function of $\left(1/\tilde x - 1/x_c\right)$, 
\begin{equation}
    \frac{\sigma}{\eta}\frac{d\eta}{d\sigma} = \frac{\sigma}{\eta}\frac{1}{(C(\phi)g(\dot\Gamma))^2}\left( \frac{-2}{f^3}\mathcal{H}\frac{df}{d\sigma} + \frac{1}{f^2}\frac{d\mathcal{H}\left(\frac{1}{\tilde x} - \frac{1}{x_c} \right)}{d\left(\frac{1}{\tilde x}- \frac{1}{x_c}\right)}\frac{d \left(\frac{1}{\tilde x} - \frac{1}{x_c}\right)}{d \sigma}\right) = 1
\end{equation}
\begin{equation}
    \frac{\sigma}{\eta}\frac{d\eta}{d\sigma} = \frac{\sigma}{\eta}\frac{1}{(C(\phi)g(\dot\Gamma))^2}\left( \frac{-2}{f^3}\mathcal{H}\frac{df}{d\sigma} + \frac{1}{f^2}\frac{d\mathcal{H}\left(\frac{1}{\tilde x} - \frac{1}{x_c} \right)}{d\left(\frac{1}{\tilde x}- \frac{1}{x_c}\right)}\frac{-1}{\tilde x^2} \frac{d \tilde x}{d \sigma}\right) = 1
\end{equation}
\begin{equation}
    \frac{\sigma}{\eta}\frac{d\eta}{d\sigma} = \frac{\sigma}{\eta}\frac{1}{(C(\phi)g(\dot\Gamma))^2}\left( \frac{-2}{f^3}\mathcal{H}\frac{df}{d\sigma} + \frac{1}{f^2}\frac{d\mathcal{H}\left(\frac{1}{\tilde x} - \frac{1}{x_c} \right)}{d\left(\frac{1}{\tilde x}- \frac{1}{x_c}\right)}\frac{-1}{\tilde x^2} \frac{C(\phi)g(\dot\Gamma)}{(\phi_0 - \phi)}\frac{d f}{d \sigma}\right) = 1
\end{equation}

\begin{equation}
    \frac{\sigma}{\eta}\frac{d\eta}{d\sigma} = \frac{\sigma}{\eta}\frac{1}{(C(\phi)g(\dot\Gamma)f(\sigma))^2}\frac{df}{d\sigma}\left( \frac{-2}{f}\mathcal{H} + \frac{d\mathcal{H}\left(\frac{1}{\tilde x} - \frac{1}{x_c} \right)}{d\left(\frac{1}{\tilde x}- \frac{1}{x_c}\right)}\frac{-1}{\tilde x^2} \frac{C(\phi)g(\dot\Gamma)}{(\phi_0 - \phi)}\right) = 1
\end{equation}

\begin{figure}[t]
\begin{center}
\includegraphics[width=0.8\linewidth]{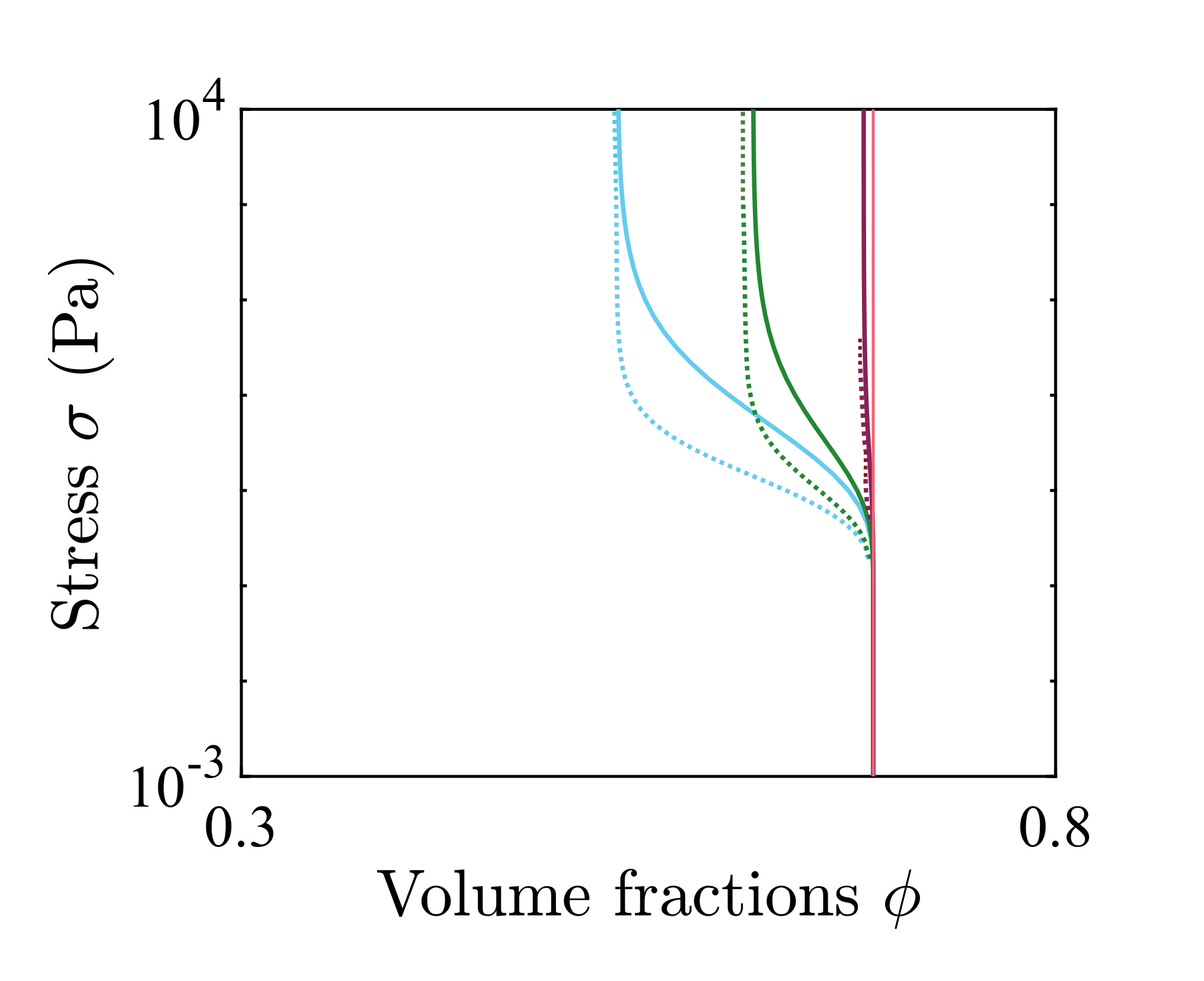}
\caption{\label{fig:PhaseDiagramWithDSTLine} \textbf{Phase diagrams with the DST Lines as estimated by using the scaling function obtained in \cite{ramaswamy2021universal}}. The DST lines are shown with the dashed lines and the shear jamming lines are the solid lines for $\dot\Gamma = 0$ (blue), $\dot\Gamma = 6.3$ (green), $\dot\Gamma = 40$ (maroon), and $\dot\Gamma = 1000$ (pink). We see that the DST region decreases as the $\dot\Gamma$ increases.}
\end{center}
\end{figure}

Substituting for the viscosity, 
\begin{equation}
 \frac{\sigma}{\frac{1}{(C(\phi)g(\dot\Gamma)f(\sigma))^2}\mathcal{H}}\frac{1}{(C(\phi)g(\dot\Gamma)f(\sigma))^2}\frac{df}{d\sigma}\left( \frac{-2}{f}\mathcal{H} + \frac{d\mathcal{H}\left(\frac{1}{\tilde x} - \frac{1}{x_c} \right)}{d\left(\frac{1}{\tilde x}- \frac{1}{x_c}\right)}\frac{-1}{\tilde x^2} \frac{C(\phi)g(\dot\Gamma)}{(\phi_0 - \phi)}\right) = 1
\end{equation}

\begin{equation}
 \sigma\frac{df}{d\sigma}\left( \frac{-2}{f} +\frac{1}{\mathcal{H}} \frac{d\mathcal{H}\left(\frac{1}{\tilde x} - \frac{1}{x_c} \right)}{d\left(\frac{1}{\tilde x}- \frac{1}{x_c}\right)}\frac{-1}{\tilde x^2} \frac{C(\phi)g(\dot\Gamma)}{(\phi_0 - \phi)}\right) = 1
\end{equation}

\begin{equation}
 \sigma\frac{df}{d\sigma}\left( \frac{-2}{f} +\frac{d \log \mathcal{H}\left(\frac{1}{\tilde x} - \frac{1}{x_c} \right)}{d\left(\frac{1}{\tilde x}- \frac{1}{x_c}\right)}\frac{-1}{\tilde x^2} \frac{C(\phi)g(\dot\Gamma)}{(\phi_0 - \phi)}\right) = 1
\end{equation}

\begin{equation}
 \label{eq:DSTSimpleEqn}
 \sigma\frac{df}{d\sigma}\left( \frac{-2}{f} - \frac{1}{\left(\frac{1}{\tilde x} - \frac{1}{x_c}\right)} \frac{d \log \mathcal{H}(t)}{d\log t}\frac{1}{\tilde x^2} \frac{C(\phi)g(\dot\Gamma)}{(\phi_0 - \phi)}\right) = 1
\end{equation}

where $t = 1/\tilde x - 1/x_c$. Using the fact that $f(\sigma) = e^{-(\sigma^*/\sigma)^{0.75}}$, 

\begin{equation}
    \frac{df}{d\sigma} = e^{-(\sigma^*/\sigma)^{0.75}} (\sigma^*)^{-0.75}(0.75)(\sigma)^{-1.75}
\end{equation}
 or 
 \begin{equation}
    \frac{df}{d\sigma} = (0.75) f \left(\frac{\sigma^*}{\sigma}\right)^{-0.75}\left(\frac{1}{\sigma}\right)
\end{equation}
 
Substituting in Eq.~\ref{eq:DSTSimpleEqn}, 

\begin{equation}
 \label{eq:DSTSimpleEqn2}
 (0.75)  \left(\frac{\sigma^*}{\sigma}\right)^{-0.75}\left( -2 - \frac{x_c}{(x_c - \tilde x)} \frac{d \log \mathcal{H}(t)}{d\log t}\right) = 1
\end{equation}
 
We can use the fit from the steady state data and substitute for $\mathcal{H}$:
\begin{equation}
\label{eq:HFit}
    \log \mathcal{H}(y) = ay + b + \left( \frac{1 + \tanh(e(y - f))}{2}\right)\left( (c-a)y + d -b\right)
\end{equation}
where $y = \log(1/\tilde x - 1/x_c)$, and $a$, $b$, $c$, $d$, $e$ and $f$ are fitting parameters. We can take the derivative of this function: 
\begin{equation}
    \label{eq:DH}
    \frac{d\log \mathcal{H}}{dy} = a + ((c - a)y + d - b)\frac{1}{2\cosh^2(e(y - f))}e + \left(\frac{1 +  \tanh(e(y - f))}{2}\right)(c - a)
\end{equation}

From Eq.~\ref{eq:DH} and Eq.~\ref{eq:DSTSimpleEqn2}, and the fits for $a$, $b$, $c$, $d$, $e$ and $f$, we can solve for the DST line at different values of $g(\dot\Gamma)$ (Fig.~\ref{fig:PhaseDiagramWithDSTLine}). We find that as the OSP flow increases, the DST area shrinks until it is nonexistent at $\dot\Gamma \rightarrow \infty$. The exact parameters used to generate the phase diagram in Fig.~\ref{fig:PhaseDiagramWithDSTLine} is shown in Table 1.

\begin{table}
\begin{center}
\begin{tabular}{ |c|c| } 
 \hline
 \textbf{Parameters} & \textbf{Values} \\ 
 \hline
 $C(\phi)$ & -3.5$\phi$ + 2.8\\ 
 \hline
 $\sigma^*$ & 1.8827\\ 
 \hline
 $\phi_0$ & 0.688\\ 
 \hline
 $x_c$ & 6.1 \\
 \hline
 Fit $\mathcal{H}$ - $a$ & -1.5 $\pm$ 0.3 \\
 \hline
 Fit $\mathcal{H}$ - $b$ & -1 $\pm$ 1 \\
 \hline
 Fit $\mathcal{H}$ - $c$ & -2.00 $\pm$ 0.04\\
 \hline
 Fit $\mathcal{H}$ - $d$ & -1.3 $\pm$ 0.3\\
 \hline
 Fit $\mathcal{H}$ - $e$ & 0.3 $\pm$ 0.2\\
 \hline
 Fit $\mathcal{H}$ - $f$ & -3 $\pm$ 2\\
 \hline
 \end{tabular}
\label{table:ParametersPhi0}
\end{center}
\caption{Parameters to determine the shear jamming and DST lines for the phase diagram.}
\end{table}

\bibliography{references}
\bibliographystyle{naturemag}